\documentclass[twocolumn]{aastex631}
\usepackage{upgreek}
\usepackage{amsmath, amstext}
\usepackage{array}
\usepackage{empheq}
\usepackage{comment}
\usepackage{mathrsfs}
\usepackage{textcomp}
\usepackage{enumitem}   
\usepackage{gensymb}
\usepackage{hyperref}
\usepackage{graphicx}
\usepackage[caption=false]{subfig}
\usepackage{multirow}
\usepackage{float}
\usepackage{longtable}
\usepackage{needspace}
\usepackage{tabularx}
\usepackage[flushleft]{threeparttable}
\usepackage{booktabs} \usepackage{CJKfntef,CJK}
\usepackage{afterpage}
\hypersetup{colorlinks, linkcolor={blue}, citecolor={blue}, urlcolor={blue}} 
\usepackage{rotating}

\definecolor{DarkOrange}{RGB}{204, 85, 0}
\definecolor{LincolnGreen}{RGB}{17, 102, 0}

\usepackage{xspace}

\newcommand{\cxo}{{Chandra}}
\newcommand{\xmm}{{XMM-Newton}}

\begin{document}
\begin{CJK*}{UTF8}{gbsn}
\pagenumbering{arabic}

\title{A new X-ray census of rotation powered pulsars}
\correspondingauthor{Shan-Shan Weng}
\email{wengss@njnu.edu.cn}

\author[0009-0000-3830-9650]{Yu-Jing Xu (徐雨婧)}\thanks{Yu-Jing Xu and Han-Long Peng contributed equally to this work.}
\affiliation{Department of Astronomy, Xiamen University, Xiamen, 361005, Fujian, People’s Republic of China}
\affiliation{School of Physics and Technology, Nanjing Normal University, Nanjing, 210023, Jiangsu, People’s Republic of China}

\author[0009-0009-8477-8744]{Han-Long Peng(彭寒龙)}\thanks{Yu-Jing Xu and Han-Long Peng contributed equally to this work.}
\affiliation{School of Physics and Technology, Nanjing Normal University, Nanjing, 210023, Jiangsu, People’s Republic of China}

\author[0000-0001-7595-1458]{Shan-Shan Weng(翁山杉)}\affiliation{School of Physics and Technology, Nanjing Normal University, Nanjing, 210023, Jiangsu, People’s Republic of China}
\affiliation{Institute of Physics Frontiers and Interdisciplinary Sciences, Nanjing Normal University, Nanjing, 210023, Jiangsu, People’s Republic of China}

\author[0000-0002-9392-547X]{Xiao Zhang(张潇)}
\affiliation{School of Physics and Technology, Nanjing Normal University, Nanjing, 210023, Jiangsu, People’s Republic of China}
\affiliation{Institute of Physics Frontiers and Interdisciplinary Sciences, Nanjing Normal University, Nanjing, 210023, Jiangsu, People’s Republic of China}

\author[0000-0002-2749-6638]{Ming-Yu Ge(葛明玉)}
\affiliation{Key Laboratory of Particle Astrophysics, Institute of High Energy Physics, Chinese Academy of Sciences, Beijing 100049, China}

\begin{abstract}

To date, over 4000 pulsars have been detected. In this study, we identify 231 X-ray counterparts of Australia Telescope National Facility (ATNF) pulsars by performing a spatial cross match across the Chandra, XMM-Newton observational catalogs. This dataset represents the largest sample of X-ray counterparts ever compiled, including 98 normal pulsars (NPs) and 133 millisecond pulsars (MSPs). Based on this significantly expanded sample, we re-establish the correlation between X-ray luminosity and spin-down power, given by $L_{\rm X} \propto \dot{E}^{0.85\pm0.05}$ across the whole X-ray band. The strong correlation is also observed in hard X-ray band, while in soft X-ray band there is no significant correlation. Furthermore, $L_{\rm X}$ shows a strong correlation with spin period and characteristic age for NPs.  For the first time, we observe a strongly positive correlation between  $L_{\rm X}$ and the light cylinder magnetic field ($B_{\rm lc}$) for MSPs, with both NPs and MSPs following the relationship $L_{\rm X} \propto B_{\rm lc}^{1.14}$, consistent with the outer-gap model of pulsars that explains the mechanism of X-ray emission. Additionally, we investigate potential X-ray counterparts for Galactic Plane Pulsar Snapshot pulsars, finding a lower likelihood of detection compared to ATNF pulsars. 

\textit{Unified Astronomy Thesaurus concepts}: Pulsars (1306); Rotation powered pulsars (1408); Neutron stars (1108)

\end{abstract}

\vspace{1em}

\section{Introduction}

In the 1930s, stellar evolution theories predicted that a neutron star (NS) could form when a massive star exhausts its fuel and undergoes gravitational collapse \citep{Baade1934}. Many years later, the existence of NSs was confirmed by the discovery of radio pulsars in 1967 August \citep{Hewish1968}. NSs can be powered either by rotational kinetic energy, magnetic energy, or accretion, and they manifest in various ways. Rotation-powered pulsars (RPPs) account for over 90\% of the population, emitting beams of electromagnetic radiation due to high-energy processes occurring at the magnetic poles or in the surrounding region \citep{Lyne2012}. 

There are currently $\sim$3630 RPPs in the Australia Telescope National Facility (ATNF) Pulsar Catalogue \citep[version 2.3.0, ][]{Manchester2005}\footnote{\protect{https://www.atnf.csiro.au/people/pulsar/psrcat/}}.  The Fermi/LAT mission has revealed that about 10\% of the known RPPs are visible in the gamma-ray band \citep{Smith2023}. Furthermore, over a hundred X-ray RPPs have been reported \citep{Becker1997, Becker1999, Chang2023}.  In contrast, the optical and IR emissions have been poorly studied \citep{Mignani2011, Mignani2018}, and optical pulsations have been revealed for less than 1\% of NSs \citep{Lyne2012, Ambrosino2017}. The radiation mechanisms of broadband emissions from RPPs are not yet completely clarified, especially for photon energies lower than the gamma-ray band. In this context, much effort has been devoted to statistical studies. A notable positive correlation between X-ray luminosity ($L_{\rm X}$) and pulsar spin-down power ($\dot{E}$) has been documented \citep[e.g. ][]{Becker1997, Rea2012, Chang2023}. However, for some pulsars, their distances still have large uncertainties, and X-ray emission may arise from a mix of components. As a result, the scatter in this relation is large \citep[e.g. ][]{Possenti2002, Li2008}, highlighting the need for more data to better understand X-ray emission from RPPs.

The Five-hundred-meter Aperture Spherical radio Telescope (FAST) is the most sensitive radio telescope used for discovering pulsars \citep{Nan2011, Jiang2019}. FAST has discovered more than 1000 pulsars \citep{Li2018, Qian2019, Weng2022, Han2021, Han2024}, particularly with two key science projects: the Commensal Radio Astronomy FAST Survey \citep{Li2018} and the Galactic Plane Pulsar Snapshot (GPPS) survey \citep{Han2021, Han2024}. Until 2024 November, the GPPS program had discovered 751 pulsars\footnote{\protect{http://zmtt.bao.ac.cn/GPPS/GPPSnewPSR.html}}, including some very faint sources. Detailed studies of these sources are progressing steadily \citep{Su2023, Zhou2023}. 

In this paper, we use all available archived X-ray observations from the Chandra and XMM-Newton missions to explore potential associations between X-ray point sources and pulsars identified in the ATNF and GPPS catalogs. 
In Section 2, we present a spatial matching analysis to construct the pulsar sample, and examine the correlation between X-ray luminosity and timing parameters for ATNF pulsars. In Section 3, we employ the results of the $L_{\rm X} - \dot{E}$ correlation to constrain possible counterparts for GPPS pulsars. Finally, a discussion and our conclusions are presented in Section 4.

\section{ATNF Pulsars}
\subsection{Sample construction}

Various mechanisms have been  proposed to explain the generation of X-ray emissions from pulsars.  For some young pulsars, surface thermal radiation (e.g., ``Magnificent Seven" and ``Central Compact Objects") or magnetic field decay (e.g., ``Rotating Radio Transient" and Magnetars) may contribute to the X-ray emission \citep{Torii1998, Gotthelf2013, Chang2023}. In certain cases,  this results in X-ray luminosity that exceeds the pulsar's spin-down luminosity. X-ray emission from Be/gamma-ray binaries is thought to arise from the shock between the stellar wind and the pulsar wind  \citep{Johnston1992, Miller2013, Weng2022}. Alternatively, in this study, we focus exclusively on RPPs where rotational energy is the dominant source of X-ray emission. The nonthermal power-law component of their emission can be interpreted as magnetospheric radiation. The origin of the thermal emission, however, remains debated and is likely due to the bombardment of charged particles returning from the magnetosphere onto the polar cap region.

As a leading international facility in radio astronomy, the ATNF catalog offers a comprehensive sample of pulsars, encompassing all published RPPs while excluding accretion-powered pulsars. In this work, we search for X-ray counterparts of 3630 pulsars listed in the ATNF pulsar catalog v2.3.0\footnote{\protect{https://www.atnf.csiro.au/people/pulsar/psrcat/}} \citep{Manchester2005} by using the XMM-Newton Serendipitous Source Catalog \footnote{\protect{https://heasarc.gsfc.nasa.gov/W3Browse/xmm-newton/xmmssc.html}} \citep[4XMM-DR13 Version; ][]{Webb2020}, the Chandra Source Catalog Release 2.0 \footnote{\protect{https://cxc.cfa.harvard.edu/csc2.1/index.html}} \citep[CSC 2.0; ][]{Evans2010, Evans2024}. If the angular separation between a pulsar and an X-ray source, with a detection confidence level greater than 3$\sigma$, satisfies the condition $\delta = \sqrt{\rm (R.A._{X} - R.A._{R})^{2}cos^{2}(decl._{X})+(decl._{X} - decl._{R})^{2}} < R_{\rm X} - R_{\rm R}$, we consider the X-ray source to be the counterpart of the pulsar. R.A. and decl. denote the right ascension and declination of a source, while $R_{\rm X}$ and $R_{\rm R}$ represent the positional uncertainties at the 2$\sigma$ confidence level for the X-ray source and the pulsar, respectively.  For most radio-loud pulsars, the timing procedure can help us to achieve positional accuracy down to milliarcseconds, and we use the Third Fermi/LAT Catalog of Gamma-ray Pulsars Catalog (3PC)\footnote{\protect{https://fermi.gsfc.nasa.gov/ssc/data/access/lat/3rd\_PSR\_catalog/}} \citep{Smith2023} to obtain more precise positions for radio-quiet and radio-faint pulsars. The typical positional accuracy of XMM-Newton serendipitous point source detections is generally less than 1.$''$57, and the on-axis spatial resolution of Chandra data is subarcsecond. However, the X-ray positional accuracy can sometimes be overestimated, leading to some counterparts being missed. We verified the results by cross-referencing with compiled literature tables, identifying 19 counterparts for eight normal pulsars (NPs) and 11 millisecond pulsars \citep[MSP; ][]{Li2008, Lee2018, Zelati2020, Chang2023}. In summary, by positional cross-matching, we identify 121 counterparts of pulsars, comprising 65 MSPs and 56 NPs.

\startlongtable
\tabletypesize{\fontsize{7.5}{8}\selectfont}
\begin{deluxetable}{cccccccccc}
\setlength{\tabcolsep}{1.5pt}
\tablecaption{X-Ray Counterpart of RPPs}
\tablehead{\colhead{Name} & \colhead{X-ray Source} & \colhead{$P$}& \colhead{$\dot{P}$}& \colhead{$D$}& \colhead{$L_{\rm X}$} & \colhead{$L_{\rm SX}$} & \colhead{$L_{\rm HX}$}& \colhead{$L_{\rm G}$}&  \colhead{Ref}  \\
 \colhead{} &  \colhead{} &  \colhead{(s)} &  \colhead{(s/s)} &  \colhead{(kpc)} &  \colhead{($\rm erg~s^{-1}$)} &  \colhead{($\rm erg~s^{-1}$)} &   \colhead{($\rm erg~s^{-1}$)} &  \colhead{($\rm erg~s^{-1}$)} &  \colhead{} }
 
\startdata
\textbf{NP} \\
J0002+6216 & 2CXO J000258.1+621609 & 0.115364 & $5.97\times10^{-15}$ & 6.357 & $1.50\times10^{32}$ & $1.08\times10^{32}$ & $3.58\times10^{31}$ & $9.01\times10^{33}$ & (T) \\
J0007+7303 & 2CXO J000701.5+730308$^{e}$ & 0.315873 & $3.60\times10^{-13}$ & 1.4 & $1.75\times10^{31}$ & $5.63\times10^{30}$ & $9.74\times10^{30}$ & $1.01\times10^{35}$ & (T, 1) \\
J0058-7218 & 2CXO J005816.8-721805$^{e}$ & 0.021766 & $2.89\times10^{-14}$ & 59.7 & $6.54\times10^{34}$ & $1.76\times10^{34}$ & $4.90\times10^{34}$ & --- & (T, 2) \\
J0108-1431 & 4XMM J010808.3-143150 & 0.807565 & $7.70\times10^{-17}$ & 0.21 & $6.99\times10^{28}$ & $4.03\times10^{28}$ & $2.79\times10^{28}$ & --- & (T) \\
\ldots & \ldots & \ldots & \ldots & \ldots & \ldots & \ldots & \ldots & \ldots & \ldots \\
J2055+2539 & 4XMM J205548.9+253958$^{e}$ & 0.319561 & $4.10\times10^{-15}$ & 0.62 & $9.89\times10^{29}$ & $4.06\times10^{29}$ & $5.82\times10^{29}$ & $2.44\times10^{33}$ & (T, 36) \\
J2139+4716$^{c}$ & 2CXO J213955.9+471613 & 0.282849 & $1.79\times10^{-15}$ & $<0.8$ & $<5.7\times10^{29}$ & $<1.1\times10^{29}$ & $<4.88\times10^{29}$ & $<4.65\times10^{35}$ & (37)\\
J2225+6535$^{b}$ & 2CXO J222552.8+653536 & 0.682542 & $9.66\times10^{-15}$ & 0.9 & $1.45\times10^{30}$ & --- & --- & --- & (3) \\
J2229+6114$^{b}$ & 2CXO J222905.2+611409$^{e}$ & 0.051648 & $7.74\times10^{-14}$ & 3 & $3.82\times10^{32}$ & --- & --- & $2.59\times10^{35}$ & (3, 4) \\
\hline
\textbf{MSP} \\
J0023+0923 & 2CXO J002316.8+092323 & 0.00305 & $1.14\times10^{-20}$ & 1.818 & $6.87\times10^{30}$ & $6.09\times10^{30}$ & $7.71\times10^{29}$ & $3.03\times10^{33}$ & (T) \\
J0024-7204C$^{ad}$ & --- & 0.005757 & $-4.99\times10^{-20}$ & 4.52 & $1.70\times10^{30}$ & --- & --- & --- & (38) \\
J0024-7204D$^{ad}$ & --- & 0.005358 & $-3.42\times10^{-21}$ & 4.52 & $3.30\times10^{30}$ & --- & --- & --- & (38) \\
J0024-7204E$^{a}$ & --- & 0.003536 & $9.85\times10^{-20}$ & 4.52 & $5.00\times10^{30}$ & --- & --- & --- & (38) \\
\ldots & \ldots & \ldots & \ldots & \ldots & \ldots & \ldots & \ldots & \ldots & \ldots \\
J2241-5236 & 4XMM J224142.0-523635$^{e}$ & 0.002187 & $6.90\times10^{-21}$ & 1.042 & $6.30\times10^{30}$ & $5.00\times10^{30}$ & $1.23\times10^{30}$ & $3.25\times10^{33}$ & (T, 4) \\
J2256-1024 & 2CXO J225656.3-102434 & 0.002295 & $1.14\times10^{-20}$ & 2.083 & $2.09\times10^{31}$ & $1.40\times10^{31}$ & $6.98\times10^{30}$ & $4.25\times10^{33}$ & (T) \\
J2302+4442 & 4XMM J230246.9+444222 & 0.005192 & $1.39\times10^{-20}$ & 0.863 & $2.75\times10^{30}$ & $2.10\times10^{30}$ & $4.85\times10^{29}$ & $3.47\times10^{33}$ & (T) \\
J2339-0533 & 4XMM J233938.7-053305 & 0.002884 & $1.41\times10^{-20}$ & 1.1 & $3.77\times10^{31}$ & $1.00\times10^{31}$ & $2.77\times10^{31}$ & $5.90\times10^{33}$ & (T) \\
\enddata
\tablecomments{ \\
$L_{\rm X}$: X-ray luminosity in 0.3--10.0 keV for \xmm, and in 0.5-7.0 keV for \cxo. \\
$L_{\rm SX}$: SX luminosity in 0.3--2.0 keV for \xmm, and in 0.5-2.0 keV for \cxo. \\
$L_{\rm HX}$: HX luminosity in 2.0--10.0 keV for \xmm, and in 2.0-7.0 keV for \cxo. \\
$L_{\rm G}$: Gamma-ray luminosity in 100 MeV. \\
$L$: The luminosity error is derived from the flux error in the catalogs, which ranges between $10\%$ and $30\%$. However, the error introduced by the DM is dominant but not included in the table. \\
\small $^{a}$: The pulsars are MSP in GCs, and their X-ray luminosities  are taken from \cite{Zhao2022}. \\
\small $^{b}$: The pulsars are associated with PWNe or SNRs, and their X-ray luminosities are obtained from the references listed in the "Ref" column.  \\
\small $^{c}$: The upper limits of these distance are inferred from the spin-down power $\dot{E}$ and the energy flux $G_{\rm 100}$ above 100 MeV, with references listed in column "Ref". The distances of other pulsars are acquired from he ATNF catalog, using the YMW16 electron distribution model \citep{Yao2017}. \\
\small $^{d}$: The $\dot{P}$  is either unavailable or negative, according to the ATNF catalog. These pulsars are excluded from the following correlation analysis.  \\
\small $^{e}$: These pulsars have been reported to exhibit X-ray pulsations. \\
\textbf{Reference:} The data source for the distance or luminosity of the pulsar.  If two references are provided, the second one refers to the study reporting the X-ray pulsations. (T) This work; (1) \cite{Marelli2012}; (2) \cite{Ho2022}; (3) \cite{Hsiang2021}; (4) \cite{Smith2023}; (5) \cite{McGowan2006}; (6) \cite{Ding2024}; (7) \cite{Ng2007}; (8) \cite{Tanashkin2022}; (9) \cite{Rigoselli2018}; (10) \cite{Danilenko2020}; (11) \cite{Hermsen2018}; (12) \cite{Rigoselli2019}; (13) \cite{Becker2004}; (14) \cite{Renaud2010}; (15) \cite{Ho2022}; (16) \cite{Park2023}; (17) \cite{Hare2021}; (18) \cite{Gotthelf2014}; (19) \cite{Chang2023}; 
(20) \cite{Rigoselli2022}; (21) \cite{Camilo2009}; (22) \cite{Van2012}; (23) \cite{Zheng2023}; (24) \cite{Marelli2014}; (25) \cite{Duvidovich2019}; (26) \cite{Lin2014}; (27) \cite{Lin2009}; (28) \cite{Pandel2012}; (29) \cite{Lu2007}; (30) \cite{Kim2020}; (31) \cite{Zyuzin2021}; (32) \cite{Zyuzin2018}; (33) \cite{Hessels2004}; (34) \cite{Razzano2023}; (35) \cite{Marelli2016}; (36) \cite{Pletsch2012}; (37) \cite{Zhao2022}; (38) \cite{Guillot2019}; (39) \cite{Archibald2010}; (40) \cite{Papitto2015}. \\
}
\tablecomments{Table \ref{tab:long} is published in its entirety in the machine-readable format. A portion is shown here for guidance regarding its form and content.}
\label{tab:long}
\end{deluxetable}

\clearpage

\begin{figure*}
\centering
\includegraphics[width=1\textwidth]{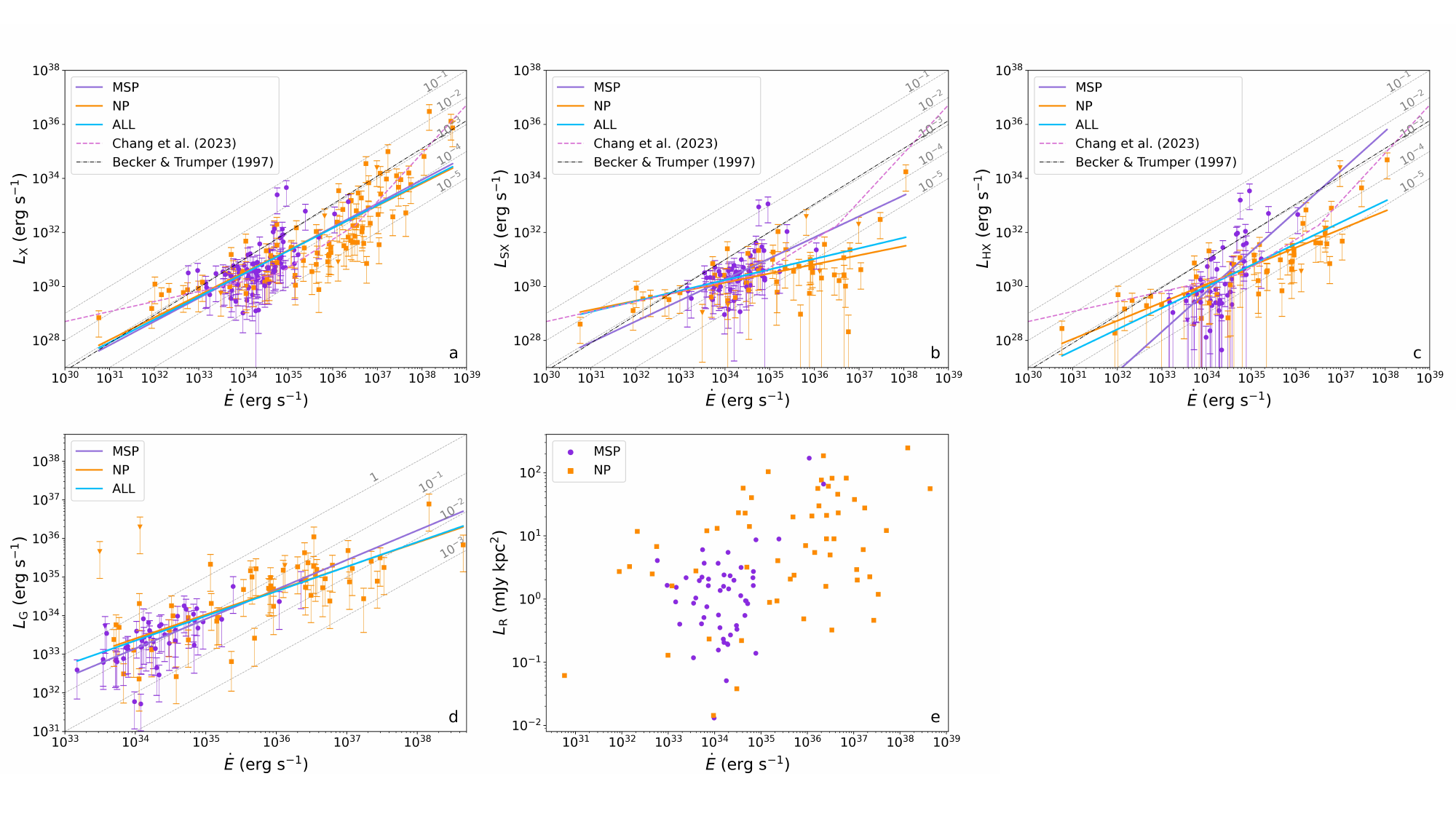}
\caption{$\dot{E}$ versus luminosities of pulsars. Panels (a)-(e) reveal correlation in X-ray band, SX band ($<$2 keV), HX band ($>$2 keV), gamma-ray band and radio band, respectively. MSPs and NPs are plotted by violet dots and orange squares, respectively. Blue, violet and orange lines are best fit of all, MSPs and NPs, respectively. The inverted triangles are upper limits of luminosity. The correlation trend is revealed by Pearson correlation coefficient ($r$), and  Spearman tests ($r_{\rm s}$ and $p_{\rm s}$), listed in Table \ref{tab:fit}. Gray dotted lines represent different values of $L_{\rm X} / \dot{E}$. The orchid dashed line and black dashed-dotted line are fitting results of \cite{Becker1997} and \cite{Chang2023}, respectively.}
\label{fig:edot}
\end{figure*}

In some cases, the positions of X-ray sources in the two X-ray catalogs are imprecise.  For sources located in supernova remnants (SNRs) or pulsar-wind nebulae (PWNe), the extended X-ray emission complicates accurate positioning. Because MSPs in globular clusters (GCs) are relatively dense and faint, they are challenging to resolve using the pipelined processing methods applied in the Chandra CSC 2.0 catalog. However, some studies have conducted more detailed analyses of these sources, yielding additional X-ray counterparts \citep[e.g. ][]{Hsiang2021, Zhao2022}. We collected data from the literature on 68 MSPs in 29 GCs and 42 NPs associated with SNRs or PWNe, incorporating these into our sample as a significant supplement. However, their luminosities are reported across the entire X-ray band because as detailed spectral analyses are lacking.

In total, we identify 231 X-ray counterparts of pulsars, including 98 NPs and 133 MSPs. Their properties, spanning radio, X-ray, and gamma-ray bands, are listed in Table \ref{tab:long}. 

\subsection{Probability of spatial coincidence}

In principle, an association can only be unambiguously confirmed as a real counterpart when X-ray pulsations are detected. However, in most cases, the time resolution of imaging X-ray observations is insufficient for pulsation searches. The time resolution of Chandra data is $\sim$3.2 s \footnote{\protect{https://cxc.harvard.edu/cdo/about\_chandra/}}, and the time resolution of XMM-Newton EPIC-pn's full frame mode observation data is 73.4 ms \footnote{\protect{https://heasarc.gsfc.nasa.gov/docs/xmm/uhb/epicmode.html}}. Nevertheless, X-ray pulsations have been reported for 51 NPs and 18 MSPs in the literature (see Table \ref{tab:long}).

We also estimate the probability of positional coincidence using the log$N$ - log$S$ distribution at the Galactic center \citep{Muno2003}. According to the discussion in Section 4.1 of \cite{Muno2003} (see also Figure 10 in their paper), the surface density of sources is extremely high, $\sim$15,000 sources deg$^{-2}$ above the flux limit of $3\times10^{-15}$ erg cm$^{-2}$ s$^{-1}$ in the 2.0--8.0 keV range. This corresponds to $\sim$0.009 contaminating sources within the 1.$''$57 location error circle  (the typical XMM-Newton positional resolution). However, we argue that the chance coincidence probability is likely much lower for the following reasons: first, the fluxes of pulsars analyzed in this work are mostly greater than $3\times10^{-15}$ erg cm$^{-2}$ s$^{-1}$, with a median value of $7\times10^{-14}/9\times10^{-15}$ erg cm$^{-2}$ s$^{-1}$ for NPs and MSPs; second, the surface density at the pulsar position should be much smaller than at the Galactic center; finally, Chandra sources exhibit even smaller positional uncertainties and a lower likelihood of coincidence. Moreover,  more refined analyses have been performed on a large proportion of MSPs in GCs  \citep[e.g. ][Section 2.1]{Hsiang2021, Zhao2022} and sources detected within PWNe can also be considered secure associations.  Therefore, we conclude that the probability of positional coincidence is very low, even if there is an absence of X-ray pulsations.

\subsection{Pulsar parameters} 

According to the magnetic dipole radiation model for pulsars, rotation power is assumed to be transformed into dipole radiation loss energy. The energy loss rate $\dot{E}$ can be expressed as $\dot{E} = \frac{4 \pi^{2} I \dot{P}}{P^{3}}$, where a typical moment of $I = 10^{45}$ g cm$^{2}$ is assumed. Other parameters can be derived from the observed rotational period $P$ and its derivative $\dot{P}$, as described by the following relations: the characteristic age $\tau = \frac{P}{2 \dot{P}}$, the surface magnetic field strength $B_{\rm surf} = 3.2\times10^{19} (P \dot{P})^{0.5}$ G, and the magnetic field at light cylinder $B_{\rm lc} = 2.9\times10^{8} P^{-2.5} \dot{P}^{0.5}$ G. 

In this work, distances are estimated using the dispersion measure (DM) based on the YMW16 electron distribution model \citep{Yao2017}. For radio-quiet or radio-faint gamma-ray pulsars, we adopt pseudodistances inferred from the spin-down power $\dot{E}$ and the energy flux $G_{\rm 100}$ above 100 MeV. As a result, upper limits on luminosities across all bands are provided for these sources, with relevant references listed in Table \ref{tab:long}. The luminosities of radio-loud pulsars in the radio and gamma-ray bands are taken from the ATNF catalog and Fermi Pulsar catalog. However, it is important to note that the uncertainty in luminosity is typically dominated by the uncertainty in the DM distance, and could deviate significantly from the actual value. For example, the distance of 93 pc for PSR J1057-5226 is derived using the YMW16 model \citep{Yao2017}, while NE2001 model places it at 720 pc \citep{Cordes2002}, showing an order-of-magnitude difference \citep{Kerr2018}. Moreover, it has been cautioned that the DM distances for some pulsars are overestimated and questionable, especially for the pulsars in the direction of the Local Arm or the tangential direction of spiral arms \citep{Han2021}.  Following \cite{Chang2023}, we adopt an uncertainty of 40\% in distances in this work. The X-ray luminosity is derived from flux and distance, and its error is calculated based on the error transfer formula.

\subsection{Correlation of parameters} 

It has been reported that, the X-ray luminosity of RPPs is strongly correlated with $\dot{E}$, but is orders of magnitude lower than $\dot{E}$ \citep[$L_{\rm X}/\dot{E} \sim 10^{-6} - 10^{-1}$, ][]{Arzoumanian2011, Rea2012, Vahdat2022, Chang2023}. The correlation was first explored in the soft X-ray (SX) band (0.1--2.4 keV) by \cite{Becker1997} using ROSAT data, leading to the relation $L_{\rm X} \propto 10^{-3}\dot{E}$. \cite{Possenti2002} later extended this analysis to the 2-10 keV band, obtaining $L_{\rm X} \propto \dot{E}^{1.34}$. \cite{Li2008} made a significant advancement by separating the X-ray luminosity contributions from pulsars and their associated PWNe, reporting $L_{\rm X,psr} \propto \dot{E}^{0.92\pm 0.04}$ and $L_{\rm X,pwn} \propto \dot{E}^{1.45\pm 0.08}$ in 2-10 keV. Recently, \cite{Chang2023} used the nonthermal X-ray luminosity in 0.5--8 keV band for 68 RPPs, finding $L_{\rm X} \propto \dot{E}^{0.88\pm 0.06}$. 

In this work, we use X-ray luminosities in 0.3--10.0 keV range for XMM-Newton and in 0.5--7.0 keV range for Chandra, further categorizing the data into SX band ($<$ 2 keV), hard X-ray (HX) band ($>$2 keV). To avoid unscientific results, we exclude pulsars with unavailable or negative values for $\dot{P}$ from the analysis. The upper limits of luminosities for radio-quite or radio-faint gamma-ray pulsars are shown in Figures 1-3, but are not included in the correlation analysis. We plot $\dot{E}$ versus luminosities in the X-ray, gamma-ray and radio bands in Figure \ref{fig:edot}, employing Pearson and Spearman correlation tests to assess the strength of linear and monotonic correlations, respectively. The results, summarized in Table \ref{tab:fit}, show that the X-ray and gamma-ray bands exhibit a strong correlation between $\dot{E}$ and luminosity, approximating $L_{\rm X} \propto 2\times10^{-4}~\dot{E}$. The explicit linear fit across the entire band yields $L_{\rm X} \propto \dot{E}^{0.85\pm0.05}$, consistent with the findings of \cite{Chang2023}. The data align with the broken power-law fitting results in their study.

In general, the thermal component primarily contributes to the SXs, while the nonthermal component dominates HXs and could also contribute significantly to the SXs. Consequently, the correlation in the soft X-ray band becomes noticeably weaker (the Pearson correlation coefficient $r$ = 0.51) compared to that of \cite{Becker1997} due to the mixture of thermal and nonthermal emissions (Figure \ref{fig:edot} and Table \ref{tab:fit}). The correlation in the HX band remains strong ($r$ = 0.73), suggesting that harder X-ray emission may provide valuable insights into the process of rotational energy loss being converted into X-rays \citep{Possenti2002}.

\begin{figure*}
\centering
\includegraphics[width=0.85\textwidth]{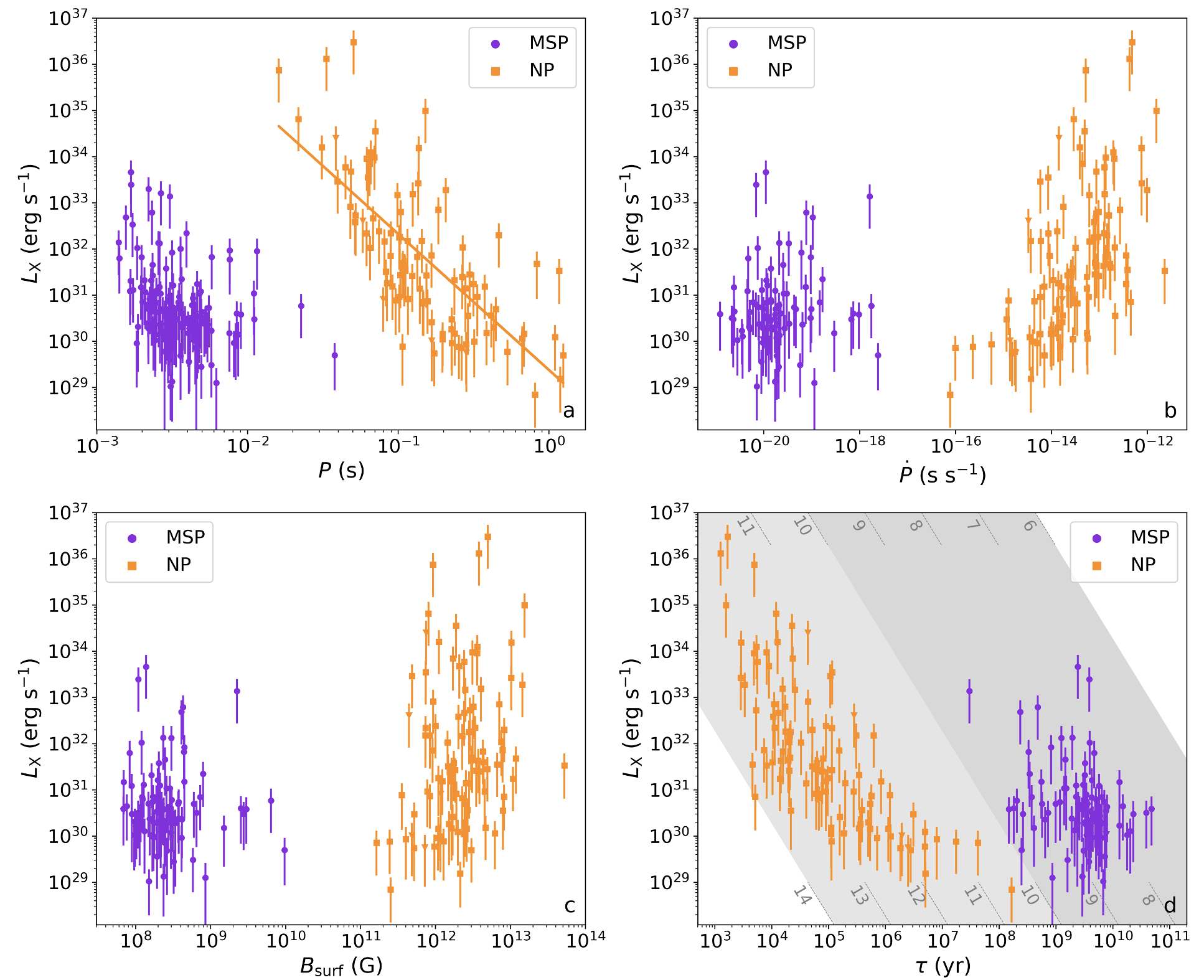}
\caption{X-ray luminosity versus timing parameters. Panels (a)-(c) reveal correlation between $L_{\rm X}$ with $P$, $\dot{P}$ and $B_{\rm surf}$. Panel (d) is a correlation between X-ray luminosity and $\tau$. The gray regions are theoretical lines for $ log B_{\rm surf} = 6, 7, 8, ..., 14$ G under the precondition that $L_{\rm X} \propto \dot{E}^{0.85}$. Dots and legends are the same as Figure \ref{fig:edot}. Linear fitting lines are plotted for Pearson correlation coefficient $r >$ 0.6.
\label{fig:lx}}
\end{figure*}

\begin{figure*}
\centering
\includegraphics[width=0.85\textwidth]{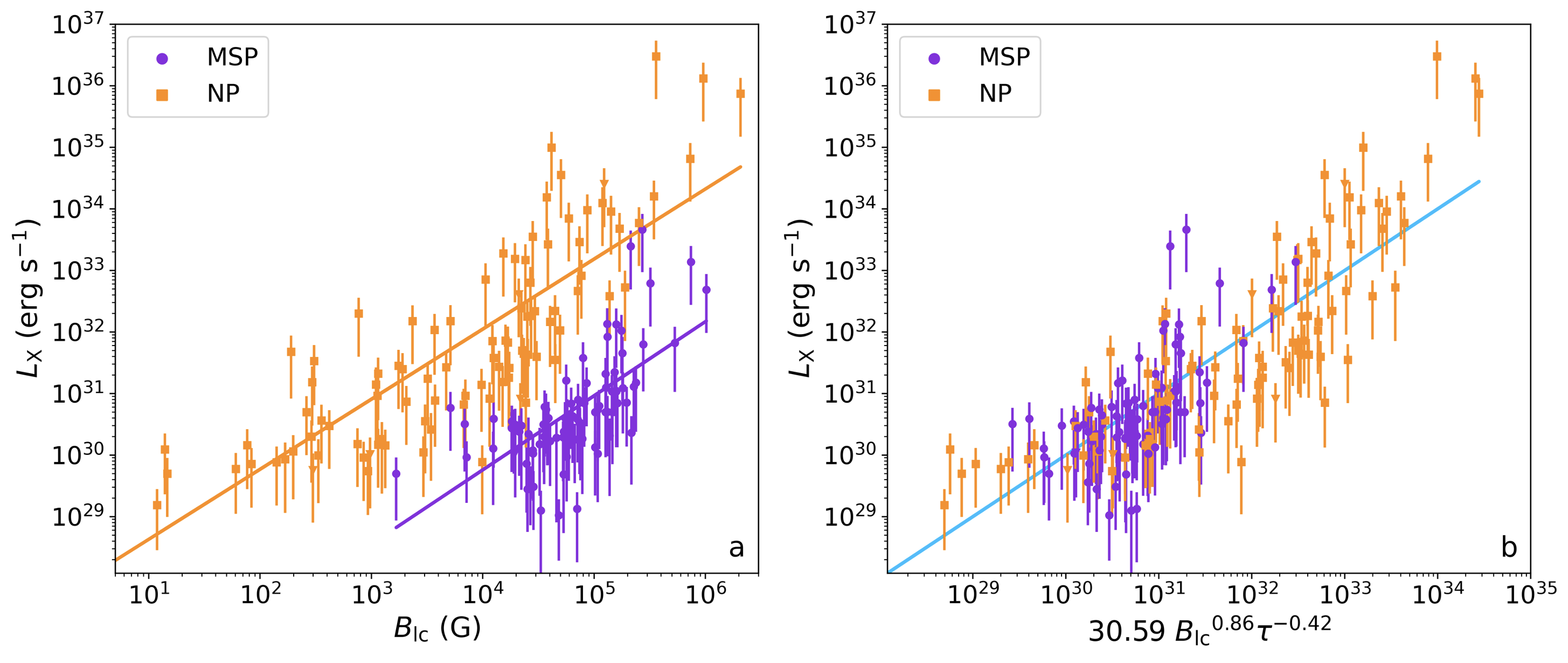}
\caption{X-ray luminosity versus $B_{\rm lc}$ and $\tau$. Dots and legends are the same as Figure \ref{fig:edot}. Panel (a) is correlation between X-ray luminosity and $B_{\rm lc}$.  Panel (b) is correlation with best fit of $B_{\rm lc}$ and $\tau$ in two dimensions. 
\label{fig:blc}}
\end{figure*}

\label{tab:fit}

\begin{table*}
\caption{Results of linear fitting and Spearman tests}

\centering
\begin{tabular}{c c  c c c c c  c c c c c  c c c c }
\hline
\multicolumn{13}{c}{a. Correlation Between $\dot{E}$ and Luminosities in Different Bands}\\
\hline
$L_{\rm X}$ & & \multicolumn{4}{c}{ALL} & & \multicolumn{4}{c}{MSP} & & \multicolumn{4}{c}{NP}  \\
\cline{3-6} \cline{8-11} \cline{13-16}
 & & $\alpha$ & $r$ & $r_{\rm s}$ & $p_{\rm s}$ & & $\alpha$ & $r$ & $r_{\rm s}$ & $p_{\rm s}$ & & $\alpha$ & $r$ & $r_{\rm s}$ & $p_{\rm s}$ \\
\hline
$L_{\rm X}$ & & $0.85\pm0.05$ & 0.83  & 0.83  & $1.18\times10^{-46}$ & & $0.87\pm0.17$ & 0.62  & 0.64  & $7.8\times10^{-11}$ & & $0.83\pm0.06$ & 0.83  & 0.86  & $6.2\times10^{-29}$ \\
 & & \multicolumn{4}{c}{(176 sources)} & & \multicolumn{4}{c}{(83 sources)} & & \multicolumn{4}{c}{(93 sources)}  \\
$L_{\rm SX}$ & & $0.39\pm0.08$ & 0.51  & 0.55  & $1.8\times10^{-10}$ & & $0.78\pm0.24$ & 0.56  & 0.57  & $2.7\times10^{-06}$ & & $0.34\pm0.09$ & 0.54  & 0.47  & $4.4\times10^{-04}$ \\
 & & \multicolumn{4}{c}{(110 sources)} & & \multicolumn{4}{c}{(58 sources)} & & \multicolumn{4}{c}{(52 sources)}  \\
$L_{\rm HX}$ & & $0.79\pm0.08$ & 0.73  & 0.75  & $1.6\times10^{-21}$ & & $1.48\pm0.25$ & 0.65  & 0.61  & $1.4\times10^{-06}$ & & $0.68\pm0.09$ & 0.83  & 0.85  & $1.5\times10^{-15}$ \\
 & & \multicolumn{4}{c}{(114 sources)} & & \multicolumn{4}{c}{(53 sources)} & & \multicolumn{4}{c}{(51 sources)}  \\
$L_{\rm G}$ & & $0.64\pm0.08$ & 0.77  & 0.76  & $2.7\times10^{-22}$ & & $0.77\pm0.24$ & 0.72  & 0.75  & $3.8\times10^{-10}$ & & $0.62\pm0.11$ & 0.79  & 0.76  & $7.0\times10^{-12}$ \\
 & & \multicolumn{4}{c}{(106 sources)} & & \multicolumn{4}{c}{(50 sources)} & & \multicolumn{4}{c}{(56 sources)}  \\
$L_{\rm R}$ & & $0.30\pm0.05$ & 0.49  & 0.46  & $4.9\times10^{-07}$ & & $0.38\pm0.14$ & 0.37  & 0.15  & $3.0\times10^{-01}$ & & $0.22\pm0.07$ & 0.40  & 0.33  & $1.2\times10^{-02}$ \\
 & & \multicolumn{4}{c}{(110 sources)} & & \multicolumn{4}{c}{(52 sources)} & & \multicolumn{4}{c}{(58 sources)}  \\
\hline
\end{tabular}
\centering
\begin{tabular}{c c  c c c c c  c c c c  }
\multicolumn{11}{c}{b. Correlation Between X-ray Luminosity and Timing Parameters} \\
\hline

$L_{\rm X}$ & & \multicolumn{4}{c}{MSP} & & \multicolumn{4}{c}{NP}  \\
\cline{3-6} \cline{8-11}
 & & $\alpha$ & $r$ & $r_{\rm s}$ & $p_{\rm s}$ & & $\alpha$ & $r$ & $r_{\rm s}$ & $p_{\rm s}$  \\
\hline
$P$ & & $-1.26\pm0.39$& -0.33& -0.36& $1.8\times10^{-05}$ & & $-2.95\pm0.27$ & -0.72  & -0.73  & $1.2\times10^{-16}$ \\
 & & \multicolumn{4}{c}{(132 sources)} & & \multicolumn{4}{c}{(93 sources)}  \\
$\dot{P}$ & & $0.13\pm0.16$& 0.10& 0.09& $4.0\times10^{-01}$ & & $1.08\pm0.12$ & 0.58  & 0.60  & $2.0\times10^{-10}$ \\
 & & \multicolumn{4}{c}{(83 sources)} & & \multicolumn{4}{c}{(93 sources)}  \\
$B_{\rm surf}$ & & $-0.10\pm0.26$& -0.05& -0.05& $6.7\times10^{-01}$ & & $0.96\pm0.24$ & 0.26  & 0.27  & $8.8\times10^{-03}$ \\
 & & \multicolumn{4}{c}{(83 sources)} & & \multicolumn{4}{c}{(93 sources)}  \\
$\tau$ & & $-0.47\pm0.19$ & -0.30  & -0.31  & $4.0\times10^{-03}$ & & $-1.18\pm0.10$ & -0.76  & -0.78  & $2.2\times10^{-20}$ \\
 & & \multicolumn{4}{c}{(83 sources)} & & \multicolumn{4}{c}{(93 sources)}  \\
$B_{\rm lc}$ & & $1.20\pm0.23$& 0.64  & 0.69& $5.0\times10^{-13}$ & & $1.14\pm0.09$ & 0.82  & 0.85  & $1.8\times10^{-27}$ \\
 & & \multicolumn{4}{c}{(83 sources)} & & \multicolumn{4}{c}{(93 sources)}  \\
\hline

\multicolumn{11}{c}{c. Correlation Between Gamma-ray Luminosity and Timing Parameters} \\
\hline
$L_{\rm G}$  & & \multicolumn{4}{c}{MSP} & \multicolumn{4}{c}{NP}  \\
\cline{3-6} \cline{8-11}
 & & $\alpha$ & $r$ & $r_{\rm s}$ & $p_{\rm s}$ & & $\alpha$ & $r$ & $r_{\rm s}$ & $p_{\rm s}$  \\
\hline
$P$ & & $-2.12\pm0.53$ & -0.53  & -0.57  & $5.0\times10^{-05}$ & & $-2.22\pm0.30$ & -0.71  & -0.70  & $2.0\times10^{-09}$ \\
 & & \multicolumn{4}{c}{(52 sources)} & & \multicolumn{4}{c}{(56 sources)}  \\
$\dot{P}$ & & $0.56\pm0.20$ & 0.41  & 0.24  & $1.2\times10^{-01}$ & & $0.61\pm0.17$ & 0.44  & 0.46  & $3.8\times10^{-04}$ \\
 & & \multicolumn{4}{c}{(50 sources)} & & \multicolumn{4}{c}{(56 sources)}  \\
$B_{\rm surf}$ & & $0.50\pm0.38$ & 0.20  & 0.06  & $6.8\times10^{-01}$ & & $0.32\pm0.36$ & 0.12  & 0.17  & $2.0\times10^{-01}$ \\
 & & \multicolumn{4}{c}{(50 sources)} & & \multicolumn{4}{c}{(56 sources)}  \\
$\tau$ & & $-0.81\pm0.17$ & -0.58  & -0.55  & $1.5\times10^{-04}$ & & $-0.80\pm0.12$ & -0.66  & -0.68  & $6.5\times10^{-09}$ \\
 & & \multicolumn{4}{c}{(50 sources)} & & \multicolumn{4}{c}{(56 sources)}  \\
$B_{\rm lc}$ & & $1.06\pm0.16$ & 0.71  & 0.75  & $7.1\times10^{-09}$ & & $0.87\pm0.09$ & 0.79  & 0.79  & $5.6\times10^{-13}$ \\
 & & \multicolumn{4}{c}{(50 sources)} & & \multicolumn{4}{c}{(56 sources)}  \\
\hline
\end{tabular}

\vspace{0.25cm}
\raggedright
{\bf Note:} $\alpha$ is the index of a power function by a linear fitting and the error is given at 1$\sigma$ confidence level, such as $L_{\rm X} \propto \dot{E}^{\alpha \pm \sigma}$. $r$ is the Pearson correlation coefficient measuring the linear correlation. $r_{\rm s}$ and $p_{\rm s}$ are the Spearman correlation coefficient and significance level measuring monotone correlation.
\end{table*}

Because additional unpulsed X-ray emission may originate from young PWNe or from intrabinary shocks in MSPs \citep{Becker1997, Becker1999, Zhang2003, Chang2023}, the correlation coefficient for MSPs ($r$ = 0.62) is generally smaller than that of NPs ($r$ = 0.83), as shown in Table \ref{tab:fit}. In our sample, 110 sources (MSPs in GCs or NPs associated with PWNe) are adopted from the literature, which only provided the luminosities in the whole X-ray band \citep{Hsiang2021, Zhao2022}.  Fitting across the entire X-ray band yields a best-fit slope ($\alpha = 0.85\pm0.05$) higher than those in the SX ($\alpha = 0.39\pm0.08$) and HX ($\alpha = 0.79\pm0.08$) bands (Figure \ref{fig:edot} and Table \ref{tab:fit}), indicating a discrepancy between the sources from literature and those derived from spatial matching. However, we cannot determine whether this small discrepancy is intrinsic or due to systematic differences in flux calculations.  The numbers of pulsars involved in each fitting are also listed in Table \ref{tab:fit}.

The linear correlation between X-ray luminosity and spin-down power strongly supports the magnetic dipole model, in which magnetic dipole radiation extracts rotational kinetic energy and causes the spin-down. We further investigate the correlation between X-ray luminosity and five timing parameters, as shown in Figures \ref{fig:lx} and \ref{fig:blc}, and Table \ref{tab:fit}. For correlations where the Pearson correlation coefficients $r >$ 0.6, we use \texttt{curve\_fit} module in \texttt{Scipy} package to plot fitting lines showing the trend, and provide fitting errors at 1$\sigma$ confidence level. In Figure \ref{fig:lx} panels (a) and (b), the best fits for $P$ and $\dot{P}$ of NPs are $L_{\rm X}$ $\propto {P}^{-2.95\pm0.27}$ and $L_{\rm X}$ $\propto \dot{P}^{1.08\pm0.12}$, respectively. Since pulsar timing parameters are functions of $P$ and $\dot{P}$, the best fit in two dimensions is found to be $L_{\rm X}$ $\propto P^{-2.55\pm0.16} \dot{P}^{0.85\pm0.04}$. In panel (c), the surface magnetic strength ($B_{\rm surf}$) shows no significant correlation with $L_{\rm X}$, consistent with the result from \citet{Chang2023}. In panel (d), the gray regions represent theoretical lines for $ {\rm log}~B_{\rm surf} = 6, 7, 8, ..., 14$ G under the assumption that $L_{\rm X} \propto \dot{E}^{0.85}$. The distribution ranges for MSPs and NPs fall between 6-10 and 10-14, consistent with the values calculated from detected $P$ and $\dot{P}$. It is worth noting that MSPs are old neutron stars “recycled” by the accretion of mass and angular momentum from a companion star in a mass-transfer binary \citep{Bhattacharya1991}, so the characteristic age $\tau$ might deviate from the real age for MSPs.

In Figure \ref{fig:blc} panel (a), we report for the first time a strong correlation between $B_{\rm lc}$ and $L_{\rm X}$ for MSPs. The fitting line reveals a consistent relationship for both NPs and MSPs, approximately $L_{\rm X} \propto B_{\rm lc}^{1.14}$. This strong correlation is also found in the gamma-ray band, with Pearson correlation coefficients $r$ = 0.71 for MSPs and $r$ = 0.79 for NPs (Table \ref{tab:fit}(c)). This result provides valuable constraints for high-energy emission models of pulsars. The outer-gap model offers an explanation for these emissions, suggesting that gamma-rays are produced in the outer-gap by electron-positron pairs ($e^{\pm}$) through inverse Compton scattering or curvature radiation processes \citep{Cheng1986}. As these gamma-rays travel back toward the neutron star surface, they convert into secondary electron-positron pairs via photon-photon pair creation. These secondary pairs then emit nonthermal X-rays through synchrotron radiation near the light cylinder \citep{Cheng1998, Takata2012}.

We also note that with the same $B_{\rm lc}$, X-ray luminosity of NPs tends to exceed that of MSPs. Therefore, two-dimensional fitting is also performed, as shown in panel (b) of Figure \ref{fig:blc}. The best fitting between $L_{\rm X}$ and $B_{\rm lc}$, $\tau$ is $L_{\rm X}$ $\propto B_{\rm lc}^{0.86\pm0.06} \tau^{-0.42\pm0.03}$. This result is consistent with the $L_{\rm X}-\dot{E}$ correlation, as $L_{\rm X} \propto B_{\rm lc}~\tau^{-0.5} \propto P^{-3} \dot{P} \propto \dot{E}$.

\section{GPPS Pulsars} 

\subsection{Sample construction and data collection} 

To discover pulsars within the Galactic latitude of $\pm10^{\circ}$ from the Galactic plane, the GPPS survey team designed the snapshot observation mode. In this mode, a sky patch of approximately 0.1575 square degrees is surveyed by four pointings using three-beam switching of the 19 beams from the L-band 19-beam receiver on FAST \citep[see ][ for more details]{Han2021}. The L-band resolution is about $2.'9$ \citep{Jiang2019}, so the initial position of a pulsar detected from one beam has an accuracy of $\leq$ 1.$'$5. This accuracy can be significantly improved ($\leq 0.''1$) with a long-term timing campaign \citep{Su2023}. We search for candidate X-ray counterparts within a circle centered on the position of GPPS pulsars with a radius of 1.$'$5. Only point-like X-ray sources are selected by setting the extent flag $EP\_Extent = 0$ for XMM-Newton sources and $extent\_flag =$ FALSE for Chandra sources. The spatial match between the GPPS pulsars and the XMM-Newton and Chandra catalogs yields 48 associations (Table \ref{tab:gpps}).

\begin{table*}
\caption{Possible X-ray counterparts of GPPS pulsars}
\centering
\label{tab:gpps}
\footnotesize
\begin{tabular}{cccc|ccccc}
\hline
Pulsar$^{a}$  & Gpps No. & Period$^{b}$ (s) & $D_{\rm YMW16}$ (kpc) & X-ray sources & R.A. & Decl. & $L_{\rm X}$ ($\rm erg~s^{-1}$) & $Var_{Flag}$\\
\hline
& & & &  4XMM &  & &  \\
J2022+3845g & gpps0076 & 1.0089 & 17.2  & J202205.4+384518 & 20:22:05.46 & +38:45:18.83 & $9.07\times10^{32}$ & FALSE \\
 &  &  &  & J202211.1+384423 & 20:22:11.19 &  +38:44:23.53 & $6.30\times10^{32}$ & FALSE \\
J2021+4024g & gpps0087 & 0.37054 & 25.0  &  J202112.9+402403 & 20:21:12.94 & +40:24:03.61 & $6.22\times10^{32}$ & FALSE \\
 &  &  &  & J202114.3+402319 & 20:21:14.35 & +40:23:19.57 & $1.27\times10^{33}$ & FALSE \\
 &  &  &  &  J202118.8+402431 & 20:21:18.83 & +40:24:31.82 & $4.93\times10^{32}$ & FALSE \\
J1852-0002g & gpps0098 & 0.2451 & 5.6  &  J185204.5-000155 & 18:52:04.48 & -00:01:57.00 & $1.90\times10^{32}$ & FALSE \\
J1907+0709g & gpps0120 & 0.3441 & 5.4  & J190756.2+070832 & 19:07:56.29 & +07:08:32.14 & $9.58\times10^{31}$ & FALSE \\
J1913+0458g & gpps0222 & 0.44479 & 4.1  &  J191337.0+045826 & 19:13:37.05 & +04:58:26.06 & $4.86\times10^{32}$ & FALSE \\
J2024+3751g & gpps0256 & 0.21164 & 15.4  &  J202429.1+374953 & 20:24:29.19 & +37:49:53.82 & $3.41\times10^{32}$ & FALSE \\
J1911+0906g & gpps0285 & 16.9259 & 1.1  &  J191135.8+090724 & 19:11:35.86 & +09:07:24.25 & $3.60\times10^{29}$ & FALSE \\
J1912+1000g & gpps0321 & 3.0528 & 4.1  &  J191244.9+095954 & 19:12:44.90 & +09:59:54.67 & $2.17\times10^{31}$ & FALSE \\
J1843-0127g & gpps0363 & 2.16489 & 7.2  &  J184332.7-012851 & 18:43:32.75 & -01:28:51.35 & $3.48\times10^{32}$ & FALSE \\
J1852-0834g & gpps0378 & 0.249315 & 6.7  &  J185218.9-083500 & 18:52:19.00 & -08:35:00.21 & $5.84\times10^{32}$ & TRUE \\
J1913+0453g & gpps0400 & 0.006086 & 15.0  &  J191346.7+045151 & 19:13:46.78 & +04:51:52.09 & $5.05\times10^{32}$ & FALSE \\
 &  &  &  &  J191346.7+045151 & 19:13:46.71 & +04:51:51.63 & $7.84\times10^{32}$ & FALSE \\
J1846-0252g & gpps0563 & 2.209439 & 6.4  &  J184627.1-025230 & 18:46:27.12 & -02:52:30.17 & $3.10\times10^{32}$ & FALSE \\
J1819-0050g & gpps0581 & 0.006602 & 4.5  &  J181933.9-005006 & 18:19:33.96 & -00:50:05.91 & $2.76\times10^{32}$ & FALSE \\
J1845-0254g & gpps0582 & 0.492655 & 5.8  &  J184532.8-025411 & 18:45:32.89 & -02:54:12.14 & $1.94\times10^{32}$ & FALSE \\
J2032+4055g & gpps0623 & 0.048739 & 10.1  &  J203237.2+405556 & 20:32:36.99 & +40:55:56.62 & $6.85\times10^{33}$ & FALSE \\
J1818-0051g	& gpps0666 & 2.20669 & 2.7 &  J181836.3-005225 & 18:18:36.37 & -0:52:25.74 & $5.65\times10^{31}$ & FALSE \\
J1847-0308g	& gpps0735 & 29.76927 & 3.4 & J184701.6-030753 & 18:47:01.65 & 18:47:01.65 & $6.57\times10^{31}$ & FALSE \\
J1851+0037g	& gpps0744 & 2.52373 & 5.2 & J185146.7+003533 & 18:51:46.39 & 18:51:46.39 & $2.09\times10^{33}$ & FALSE \\
\hline
& & & & 2CXO &  & &  \\
J1852+0056g & gpps0014 & 1.177793 & 7.2  &  J185215.4+005743 & 18:52:15.40 & +00:57:43.30 & $6.29\times10^{31}$ & FALSE \\
J1855+0139g & gpps0026 & 0.44414 & 5.2  &  J185512.5+013807 & 18:55:12.57 & +01:38:07.90 & $4.31\times10^{31}$ & TRUE \\
 &  &  &  &  J185518.9+013844 & 18:55:18.94 & +01:38:44.38 & $7.63\times10^{31}$ & FALSE \\
J1904+0519g & gpps0037 & 1.68053 & 2.5  &  J190403.8+052014 & 19:04:03.81 & +05:20:14.02 & $1.71\times10^{31}$ & FALSE \\
 &  &  &  &  J190404.8+052006 & 19:04:04.83 & +05:20:06.68 & $6.35\times10^{30}$ & FALSE \\
J2022+3845g & gpps0076 & 1.0089 & 17.2  &  J202205.4+384519 & 20:22:05.46 & +38:45:19.55 & $3.57\times10^{32}$ & FALSE \\
 &  &  &  &  J202209.5+384413 & 20:22:09.53 & +38:44:13.95 & $9.90\times10^{30}$ & TRUE \\
 &  &  &  &  J202209.9+384348 & 20:22:09.90 & +38:43:48.03 & $8.58\times10^{31}$ & FALSE \\
 &  &  &  &  J202210.8+384341 & 20:22:10.82 & +38:43:41.97 & $9.08\times10^{30}$ & FALSE \\
 &  &  &  &  J202211.2+384423 & 20:22:11.27 & +38:44:23.58 & $2.35\times10^{31}$ & FALSE \\
J2021+4024g & gpps0087 & 0.37054 & 25.0  &  J202106.0+402319 & 20:21:06.04 & +40:23:19.68 & $4.14\times10^{32}$ & FALSE \\
 &  &  &  &  J202111.7+402335 & 20:21:11.71 & +40:23:35.29 & $4.65\times10^{32}$ & FALSE \\
 &  &  &  &  J202112.8+402405 & 20:21:12.91 & +40:24:05.48 & $6.84\times10^{31}$ & TRUE \\
 &  &  &  &  J202114.3+402520 & 20:21:14.35 & +40:25:20.41 & $8.58\times10^{31}$ & FALSE \\
J1907+0658g & gpps0127 & 0.21834 & 7.7  &  J190737.8+065841 & 19:07:37.84 & +06:58:41.02 & $4.78\times10^{31}$ & FALSE \\
J1909+0905g & gpps0178 & 1.49488 & 5.4  &  J190935.9+090600 & 19:09:35.92 & +09:06:00.44 & $5.88\times10^{31}$ & FALSE \\
J1953+1844 & gpps0190 & 0.004441 & 4.3  &  J195337.9+184454 & 19:53:37.96 & +18:44:54.40 & $1.25\times10^{30}$ & FALSE \\
J1931+1841g & gpps0233 & 2.59411 & 5.4  &  J193111.2+183934 & 19:31:11.22 & +18:39:34.27 & $3.53\times10^{31}$ & FALSE \\
J2030+3833g & gpps0295 & --- & 15.2  &  J203024.9+383322 & 20:30:25.00 & +38:33:22.98 & $9.17\times10^{32}$ & FALSE \\
J1844-0223g & gpps0493 & 0.65772 & 6.3  &  J184516.7-022929 & 18:45:16.78 & -02:29:29.63 & $3.63\times10^{32}$ & FALSE \\
J1929+1337g & gpps0495 & 0.203318 & 7.8  &  J192924.8+133637 & 19:29:24.83 & +13:36:37.19 & $3.73\times10^{32}$ & FALSE \\
J1915+1045g & gpps0518 & 1.54588 & 3.7  &  J191531.5+104333 & 19:15:31.54 & +10:43:33.80 & $1.03\times10^{30}$ & FALSE \\
J2032+4055g & gpps0623 & 0.048739 & 10.1  &  J203234.8+405617 & 20:32:34.87 & +40:56:17.23 & $1.76\times10^{32}$ & TRUE \\
 &  &  &  &  J203236.3+405529 & 20:32:36.33 & +40:55:29.62 & $2.23\times10^{32}$ & FALSE \\
 &  &  &  &  J203240.2+405348 & 20:32:40.21 & +40:53:48.85 & $2.56\times10^{31}$ & FALSE \\
J1843-0310g & gpps0672 & 0.285151 & 8.5 &  J184305.3-030954 & 18:43:05.31 & -03:09:54.86 & $2.92\times10^{32}$ & FALSE \\
\hline
\multicolumn{9}{l}{\small $^{a}$: Pulsar names with a suffix `g' indicate the temporary nature, due to position uncertainty of about 1$'$5. } \\
\multicolumn{9}{l}{\small $^{b}$: Several pulsars are discovered due to single pulses and their spin periods are currently unavailable \citep{Zhou2023}.} \\
\multicolumn{9}{l}{\small  $D_{\rm YMW16}$: Distance estimated based on the YMW16 electron distribution model \citep{Yao2017}.}\\
\multicolumn{9}{l}{\small  $Var_{Flag}$: The flag is set to `True' if the source displayed flux variability within one or between observations }\\
\multicolumn{9}{l}{or to `False' if the source was tested for variability but did not qualify.}

\end{tabular}

\end{table*}

\subsection{Parameters analysis}

\begin{figure}
\centering
\includegraphics[width=0.5\textwidth]{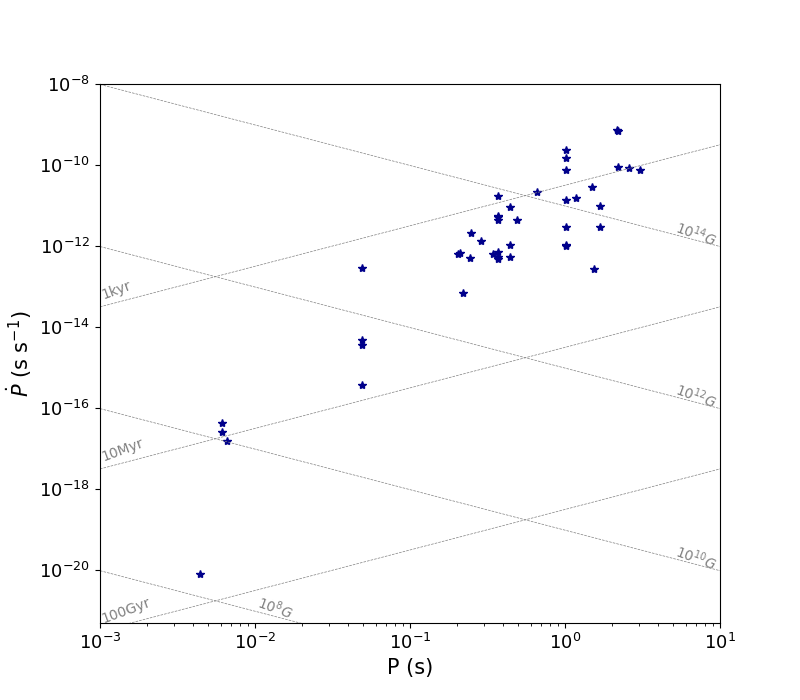}
\caption{$P-\dot{P}$  diagrams. The values of $\dot{P}$ are derived by the relationship that $L_{\rm X} \propto \dot{E}^{0.85}$.
\label{fig:ppdot}}
\end{figure}

We convert the fluxes listed in the catalogs to the isotropic X-ray luminosities in the 0.3--10 keV, 0.5--7 keV bands for XMM-Newton and Chandra catalogs, respectively. The X-ray point sources exhibit X-ray luminosities ranging from $3.6 \times 10^{29}$ erg~s$^{-1}$ to $6.8 \times 10^{33}$ erg~s$^{-1}$. Assuming that the GPPS pulars share the same $L_{\rm X} \propto \dot{E}^{0.85}$ (Figure \ref{fig:edot}) correlation with the ATNF pulsars, the spin-down power can be calculated from the observed X-ray luminosity. The period $P$ of GPPS pulsars is provided in the catalog, and the period derivative $\dot{P}$ can be derived using the formula $\dot{P} = \frac{\dot{E} P^{3}}{4 \pi^{2} I}$. For most sources, the $\dot{E}$ and $\dot{P}$ derived from the $L_{\rm X} - \dot{E}$ relation appear excessively high. It is illogical for many sources to have $\tau$ values smaller than 1 kyr or magnetic fields greater than the critical value of $4.4 \times 10^{13}$ G (Figure \ref{fig:ppdot}). Excluding these suspicious sources, there remain 27 X-ray point sources located within 16 pulsars' positional error circles. However, it is important to note that a large proportion of these X-ray point sources are not actual X-ray counterparts of the GPPS pulsars. Due to FAST's angular resolution of 2.$'$9, the probability to find an unrelated X-ray source within a positional error circle is $10^{3}-10^{4}$ times higher than that for the ATNF pulsars. Hence, there may be multiple X-ray sources surrounding the GPPS pulsars within 1.$'$5 (e.g. J1855+0139g and J2021+4024g). 

As discussed above, X-ray luminosity of RPPs is positively correlated with their rotational energy loss \citep{Becker1997, Li2008, Arzoumanian2011, Vink2011, Rea2012, Vahdat2022, Chang2023}. Assuming $L_{\rm X} \sim 2\times10^{-4}~\dot{E}$ (Figure \ref{fig:edot}), the derived values of $\dot{E}$ are larger than $10^{34}~{\rm erg~s^{-1}}$ when $L_{\rm X} > 2\times10^{30}~{\rm erg~s^{-1}}$. In contrast, more than 60 newly discovered pulsars by FAST have timing solutions, and their spin-down powers are mostly less than $10^{34}~{\rm erg~s^{-1}}$ \citep{Li2018, Su2023, Wu2023}. We estimate the fluxes of these pulsars using the detected $P$ and $\dot{P}$, finding that almost 87\% are lower than $10^{-15}~{\rm erg~cm^{-2}~s^{-1}}$. However, most sources detected by XMM-Newton and Chandra have fluxes over $10^{-14}~{\rm erg~cm^{-2}~s^{-1}}$ and $10^{-15}~{\rm erg~cm^{-2}~s^{-1}}$, respectively \citep{Webb2020, Primini2011}. Only the MSP, PSR~J1953+1844, is found to be associated with a faint X-ray point source, 2CXO~J195337.9+184454 (R.A. = 19:53:38.0, Dec. = +18:44:54.4) with a separation of less than 0.$''$2 \citep{Pan2023}. Thus we conclude that the X-ray counterparts of FAST newly discovered pulsars are less likely to be found in the mentioned XMM-Newton and Chandra catalogs compared to ATNF pulsars, and the point sources with high X-ray luminosity are likely to be coincidences with GPPS pulsars.

\section{Discussion and Summary} \label{sec:discuss}

In this work, we search for the X-ray counterparts of ATNF pulsars by cross-correlating their positions with the XMM-Newton, Chandra catalogs. A total of 231 X-ray counterparts are identified, including 98 NPs and 133 MSPs, making this the largest and most comprehensive catalog to date.

1. X-ray luminosity of pulsars shows a strong correlation with spin-down power across the entire X-ray band for both NPs and MSPs, following the relation $L_{\rm X} \propto \dot{E}^{0.85\pm0.05}$, which is consistent with the findings of \cite{Chang2023} within the error margins. The positive correlation is particularly strong in the HX band, while in the soft band the correlation weakens compared to \citet{Becker1997}, likely due to the presence of mixed thermal and nonthermal components.

2. In traditional RPP theories, magnetic dipole radiation extracts rotational kinetic energy from neutron star and transforms it into electromagnetic radiation, including X-ray emissions. This suggests an expected positive correlation between the magnetic field and electromagnetic radiation. However, no significant correlation is observed between $L_{\rm X}$ and $B_{\rm surf}$. On the other hand, we observe strong correlations between X-ray luminosity and the pulsar parameters $P$, $\tau$, and $B_{\rm lc}$ for NPs. Notably, this study is the first to report a strong correlation between $B_{\rm lc}$ and luminosities in the X-ray and gamma-ray bands of both NPs and MSPs. These findings suggest that the high-energy emission from RPPs can be more effectively explained by the outer-gap model. The best fit for $L_{\rm X}$ as a function of $B_{\rm lc}$ and $\tau$ in two dimensions is $L_{\rm X}$ $\propto B_{\rm lc}^{0.86\pm0.06} \tau^{-0.42\pm0.03}$. 

3. For all detected pulsars，luminosities in the radio and Gamma-ray bands do not show significant correlations with timing parameters. However, when examining the correlation for pulsars with X-ray counterparts listed in Table \ref{tab:long} and calculating the Pearson and Spearman correlation coefficients (see Table \ref{tab:fit}), we find that Gamma-ray luminosity shows a strong correlation with $P$, $\tau$, and $B_{\rm lc}$, similar to the correlations seen in the X-ray band. In contrast, we still find no significant correlations in the radio band.

4. We identify 27 putative associations around the 16 GPPS pulsars. However, by examining the properties of GPPS pulsars with available timing solutions, we find that their $\dot{E}$ tend to be below $10^{34}$ erg~s$^{-1}$, and most estimated fluxes are below the detection thresholds of current X-ray telescopes, unless long-term exposure is available. Consequently, we conclude that the likelihood of discovering the actual X-ray counterparts is lower compared to the ATNF pulsars. A portion of the X-ray point sources selected in this work are likely coincidental and unrelated to the GPPS pulsars.

\vspace{0.5cm}

\textit{Acknowledgements} -- 
We would like to thank the referee for their valuable suggestions and comments that improved the clarity of the paper. This research has made use of data collected by ATNF, FAST and  two X-ray missions, Chandra, and XMM-Newton.  FAST is a Chinese national mega-science facility, operated by National Astronomical Observatories, Chinese Academy of Sciences. S.S.W. thank Professors. Jin-Lin Han, Hao Tong, and Ren-Xin Xu for many valuable discussions. The authors give their thanks for support from the National Natural Science Foundation of China under Grants 12473041, 12033006, 12373051, and 12393852.

\vspace{0.5cm}

\textit{Data Availability} -- 
The XMM-Newton, and Chandra point source catalogs used in this work are available from \protect{\url{https://heasarc.gsfc.nasa.gov/W3Browse/xmm-newton/xmmssc.html}}, and \protect{\url{https://cxc.cfa.harvard.edu/csc2.1/index.html}}, respectively. The information of ATNF and FAST GPPS pulsars are available on the webpage of \protect{\url{https://www.atnf.csiro.au/people/pulsar/psrcat/}} and the GPPS survey \protect{\url{http://zmtt.bao.ac.cn/GPPS/GPPSnewPSR.html}}, respectively.

Table \ref{tab:long} will be updated regularly online: \protect{\url{https://xray-pulsar.github.io/counterparts/}}.

\vspace{0.5cm}

\facilities{CXO, XMM, FAST:500m}

\end{CJK*}

\end{document}